\documentclass[a4paper,fleqn,usenatbib]{mnras}

\usepackage[T1]{fontenc}
\usepackage{ae,aecompl}

\usepackage{graphicx}	
\usepackage{amsmath}	
\usepackage{amssymb}	

\title{The circum-galactic medium of quasars: CIV absorption systems.}

\author[M. Landoni et al]{
M. Landoni,$^{1}$\thanks{E-mail: marco.landoni@brera.inaf.it}
R. Falomo,$^{2}$
A. Treves,$^{3}$
R. Scarpa,$^{4}$
E. P.  Farina$^{5}$
\\
$^{1}$INAF - Osservatorio Astronomico di Brera, via E. Bianchi 46 Merate (LC)\\
$^{2}$INAF - Osservatorio Astronomico di Padova, vicolo dell'Osservatorio 5 Padova (PD)\\
$^{3}$Universita' degli Studi dell'Insubria, via Valleggio 11 Como (CO)\\
$^{4}$Instituto de Astrofisica de Canarias - IAC\\
$^{5}$ Max--Planck--Institut f{\"u}r
Astronomie --- K{\"o}nigstuhl 17, D-69117 Heidelberg, Germany
}

\pubyear{2015}

\begin{document}
\label{firstpage}
\pagerange{\pageref{firstpage}--\pageref{lastpage}}
\maketitle

\begin{abstract}
We investigate the properties of the circumgalactic gas in the halo of quasar host galaxies from CIV absorption line systems. Optical spectroscopy of closely aligned pairs of quasars (projected distance $\leq$ 200 kpc) obtained at the Gran Telescopio Canarias is used to investigate the distribution of the absorbing gas for a sample of 18 quasars at z $\sim$ 2. We found that the detected absorption systems of EW $\geq$ 0.3$\textrm{\AA}$ associated with the foreground QSO are revealed up to 200 kpc from the center of the host galaxy. The structure of the absorbing gas is rather patchy with a covering fraction of the gas that quickly decreases beyond 100 kpc. These results are in qualitative agreement with those found for the lower ionisation metal Mg II $\lambda$2800$\textrm{\AA}$.
\end{abstract}




\section{Introduction}
Direct observation of low redshift galaxies ($z \leq 1$) demonstrated the presence of large and diffuse warm-to-hot gas halos up to $\sim$ 200 kpc commonly referred as circumgalactic-medium (CGM), see e.g. \cite*{lanz, chen,churchill}. In the last decades, a number of papers exploited absorption lines imprinted in the spectra of background QSO to investigate the physical properties of the CGM \citep[e.g.][]{bs,churchill, mag1,mag2} finding significant correlation between the absorptions of the gas in CGM and the global properties of galaxies, such as luminosity \citep*{chen08}, mass \citep*{churchill13} , color \citep*{zibetti} and star formation rate \citep*{prochter,menard,nestor}.  
Nevertheless, only a few studies have been focused on the properties of the gaseous halo of galaxies hosting a QSO in their centre \citep[][and references therein]{hennawi,hp07}.

The standard model for the origin of the extreme luminosity of quasars (QSO) considers that a
supermassive black hole shines when intense mass inflow takes place, possibly as
a consequence of tidal forces in dissipative events \citep*{dimatteo}. In this scenario, the CGM of QSO is expected to be populated by tidal
debris, streams, and cool gas clouds, as commonly observed in interacting galaxies \citep[see e.g.][]{sulentic,cortese}. Moreover the gas of the CGM belonging to the QSO host galaxy could be metal
enriched by supernova-driven winds triggered by starbursts events associated to the mergers
or by QSO-driven outflows of gas \citep[e.g.,][]{steidel,shen}. Although in the last few years a great effort allowed to detect emission lines that arise from the CGM \citep*{hp13, martin, cantalupo,hennawi15}, only the Lyman-$\alpha$ feature has been observed so far. 
For these reasons, the most efficient way to study the CGMs is to investigate the absorption features that it imprints in the
spectra of background QSOs \citep[e.g.,][]{adel,hennawi}.

In this context projected QSO pairs are ideal observational tools for this
purpose, since the light of the very bright source in the background ($z=z_B$), goes through the
extended halo of the foreground ($z_F$ $<z_B$ ) object \citep[e.g.,][]{hennawi,farina13}.
The absorption features in the gaseous halos belonging to the foreground QSO can be therefore exploited to understand the processes of enrichment of material far from the host galaxy.
In this paper we aim to characterise the properties of intervening C IV absorbers in the circumgalactic medium of quasars host galaxies up to a projected distance (PD) of 200 kpc. We adopt the following cosmological parameters $H_{0} = 70$ km s$^{-1}$  Mpc$^{-1}$, $\Omega_{m} = 0.27$, $\Omega_{\Lambda} = 0.73$.

\section{Sample selection}
In order to investigate the properties and abundances of C IV in the CGM of quasars, we selected QSO projected pairs by searching in Sloan Digital Sky Survey DR10 of spectroscopic quasars (Paris et. al. 2012). We assume as good candidates pairs with PD (comoving transverse distance) smaller than 200 kpc in order to characterise the innermost region of the CGM. Further, we constrain the line-of-sight (LOS) velocity difference, based on the published redshift, $\Delta V \geq 5000$ km s$^{-1}$ to ensure that pairs are not gravitationally bound. We also put a threshold on the magnitude of the background QSO ($m_r \leq 20$) to secure spectra with adequate signal to noise ratio ($\geq 15$) and on declination to ensure good visibility from Roque de los Muchachos observatory. 
Finally, we selected pairs in which the redshifts of the foreground and background QSOs combine, so that the CIV absorption lines at the redshift of the foreground ($z_F$) fall within the wavelength range 4500-6000$\textrm{\AA}$. This procedures yielded 34 candidate pairs. We selected 25 objects for observation, according to the visibility of the considered period. However, only 18 targets were obtained  due to partially bad weather conditions. Details on the observed objects are reported in Table 1.

\section{Observations, Data Reduction and Data analysis}
\begin{figure*}
	\includegraphics[width=13cm]{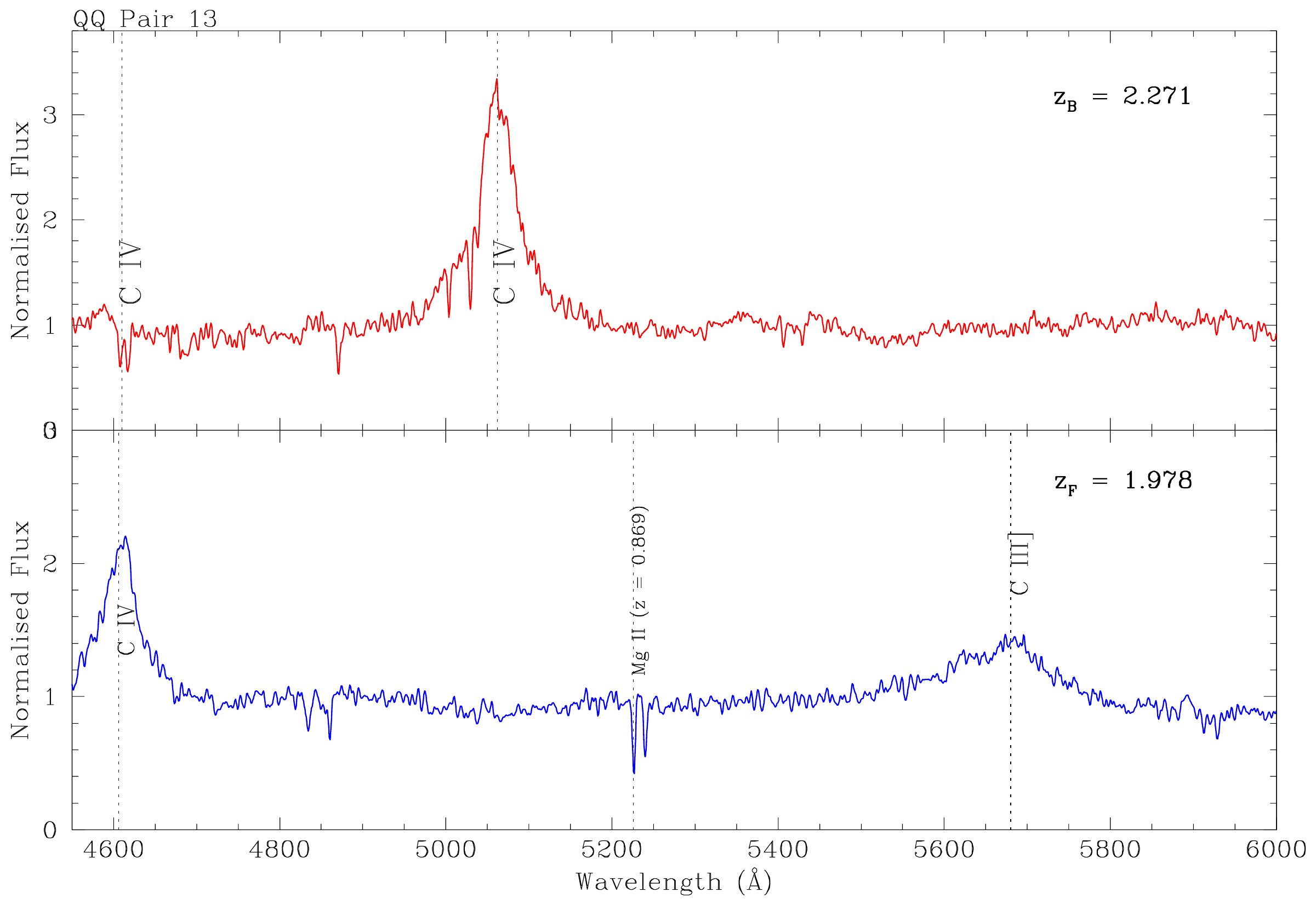}
    \caption{Normalised spectra of the pair QQ 13. Solid blue line is the spectrum of the foreground QSO while red solid line is used for the background one. The most prominent quasar emission lines are also marked. In the spectrum of the foreground we also detect an intervening Mg II absorption system at $z = 0.869$. The whole figure set is available in Electronic Version in the online edition of this Journal.}
    \label{fig:example_pair}
\end{figure*}

We observed our QSO pairs with the 10.4m Gran Telescopio Canarias (GTC) equipped with the Optical System for Imaging and Low Resolution Integrated Spectroscopy (OSIRIS, \cite*{cepa}) from September 2013 to August 2014. 
Observations were gathered with GTC-OSIRIS adopting the grism R2500V with a slit of 1.00$^{\prime\prime}$ yielding a resolution $\frac{\lambda}{\Delta\lambda} \sim 2500$ (1 px = 0.80 $\textrm{\AA}$, corresponding to 45 km s$^{-1}$ at center wavelength). In this case the resolution corresponds to a FWHM of about 2 $\textrm{\AA}$ allowing to fully resolve the components of the C IV doublets ($\lambda\lambda$1548-1551$\textrm{\AA}$). The resolution element, which corresponds to a FWHM of 120 km s$^{-1}$, is not sufficient to kinematically resolve the internal dynamics of the absorbing gas which is beyond the aim of our investigation.   
For each pair we oriented the slit in order to acquire simultaneously the spectra of the two objects and we secured three different exposures, applying a small shift of 5$^{\prime\prime}$ along the slit to better reject cosmic rays and for accounting for CCD defects. 
\\
We reduced our data by the adoption of standard $\mathtt{IRAF}$\footnote{
IRAF is distributed by the National Optical Astronomy Observatories,
which are operated by the Association of Universities for
Research in Astronomy, Inc., under cooperative agreement
with the National Science Foundation.}  procedures. Briefly, for each frame, we performed bias subtraction and flat field correction using the $\mathtt{ccdred}$ package. Wavelength calibration has been assessed through the observation of arc lamps (Xe+Ne+HgAr) and the residuals on the calibration are around 0.04$\textrm{\AA}$. We flux calibrated the spectra exploiting standard stars observed during the same nights of the targets. We corrected for systematics, slit losses and variation of the sky conditions through aperture photometry of the field, in $r^{'}$-band, acquired shortly before the observation. We report an example of spectra of a pair in Figure \ref{fig:example_pair} and we give the full figure set in the electronic edition of this Journal. 
The Galactic reddening was taken into account considering the estimates from \cite*{sch} assuming R$_V$ = 3.1 \citep*{cardelli}. In the spectra of the QSO presented in Figure 1, in addition to the typical broad emission lines of C IV and C III], we note that several absorptions lines due to intervening matter (e.g. Mg II at $z = 0.869$) are present. The C IV absorption system ascribed to the halo of the foreground QSO is also detected in some cases in the spectrum of the background one.

%
%
%
%
%
%


\begin{table*}
	\centering
	\caption{Properties of the observed pairs. Pair identification (ID), position of the foreground target (RA,DEC), foreground redshift (z$_F$), background redshift (z$_B$), V apparent magnitudes of foreground and background QSO (V$_F$, V$_B$), Projected distance in kpc (PD), Signal to noise ratios (per pixel) of foreground and background object (SN$_F$, SN$_B$). }
	\label{tab:sample}
	\begin{tabular}{llllllllll} 
		\hline
		ID & RA$_F$ & DEC$_F$ & z$_F$ & z$_B$ & V$_F$ & V$_B$ & PD & SN$_F$ & SN$_B$\\
		\hline
		QQ01  &  08:45:13.57 & +39:10:25.65  &  2.040  &  2.210  &  19.8  &  19.5  &  180  &  15  &  15\\
		QQ02  &  09:17:06.47 & +00:56:35.10  &  2.140  &  2.472  &  20.4  &  20.2  &  200  &  10  &  10\\
		QQ03  &  10:13:01.20 & +40:23:03.52  &  2.185  &  2.504  &  18.4  &  19.2  &  192  &  35  &  20\\
		QQ04  &  12:06:51.22 & +02:04:21.90  &  2.443  &  2.522  &  20.5  &  19.0  &  110  &  15  &  20\\
		QQ05  &  13:58:06.09 & +61:18:26.70  &  2.015  &  2.167  &  20.5  &  20.2  &  190  &  10  &  15\\
		QQ06  &  14:30:33.61 & -01:34:45.69  &  2.273  &  2.350  &  19.5  &  19.5  &  55  &  10  &  10\\
		QQ07  &  09:13:23.31 & +04:02:35.15  &  2.040  &  2.375  &  19.1  &  19.5  &  92  &  15  &  15\\
		QQ08  &  09:16:11.20 & -01:19:41.50  &  2.753  &  2.917  &  20.7 &  20.8 &  87  &  15  &  10\\
		QQ09  &  10:09:35.86 & +47:49:34.61  &  2.292  &  2.590  &  20.7  &  19.7  &  93  &  5  &  15\\
		QQ10  &  11:34:26.18 & +00:38:54.86  &  2.209  &  2.365  &  20.4  &  19.6  &  90  &  15  &  20\\
		QQ11  &  00:42:52.23 & +01:11:55.62  &  2.027  &  2.084  &  18.7 &  18.9 &  70  &  30  &  30\\
		QQ12  &  03:44:11.98 & +00:09:27.88  &  2.125  &  2.240  &  21.4  &  19.4  &  184  &  10  &  25\\
		QQ13  &  12:14:31.06 & +32:23:28.24  &  1.978  &  2.271  &  19.5  &  20.0  &  202  &  10  &  10\\
		QQ14  &  08:45:33.63 & +25:15:51.64  &  2.110  &  2.292  &  19.0  &  20.3 &  186  &  15  &  5\\
		QQ15  &  12:19:30.15 & +15:10:28.44  &  1.943  &  2.313  &  19.2  &  19.7  &  110  &  35  &  25\\
		QQ16  &  12:21:49.62 & +37:00:13.82  &  2.119  &  2.322  &  20.4  &  20.7  &  97  &  10  &  10\\
		QQ17  &  14:59:06.93& +12:34:49.54  &  2.109  &  2.500  &  18.6  &  20.6 &  42  &  50  &  20\\
		QQ18  &  15:53:19.30 & +31:52:40.39  &  2.817  &  3.194  &  22.3  &  20.0  &  200  &  15  &  25\\
		\hline
	\end{tabular}
\end{table*}

\begin{figure}
	\includegraphics[width=9cm]{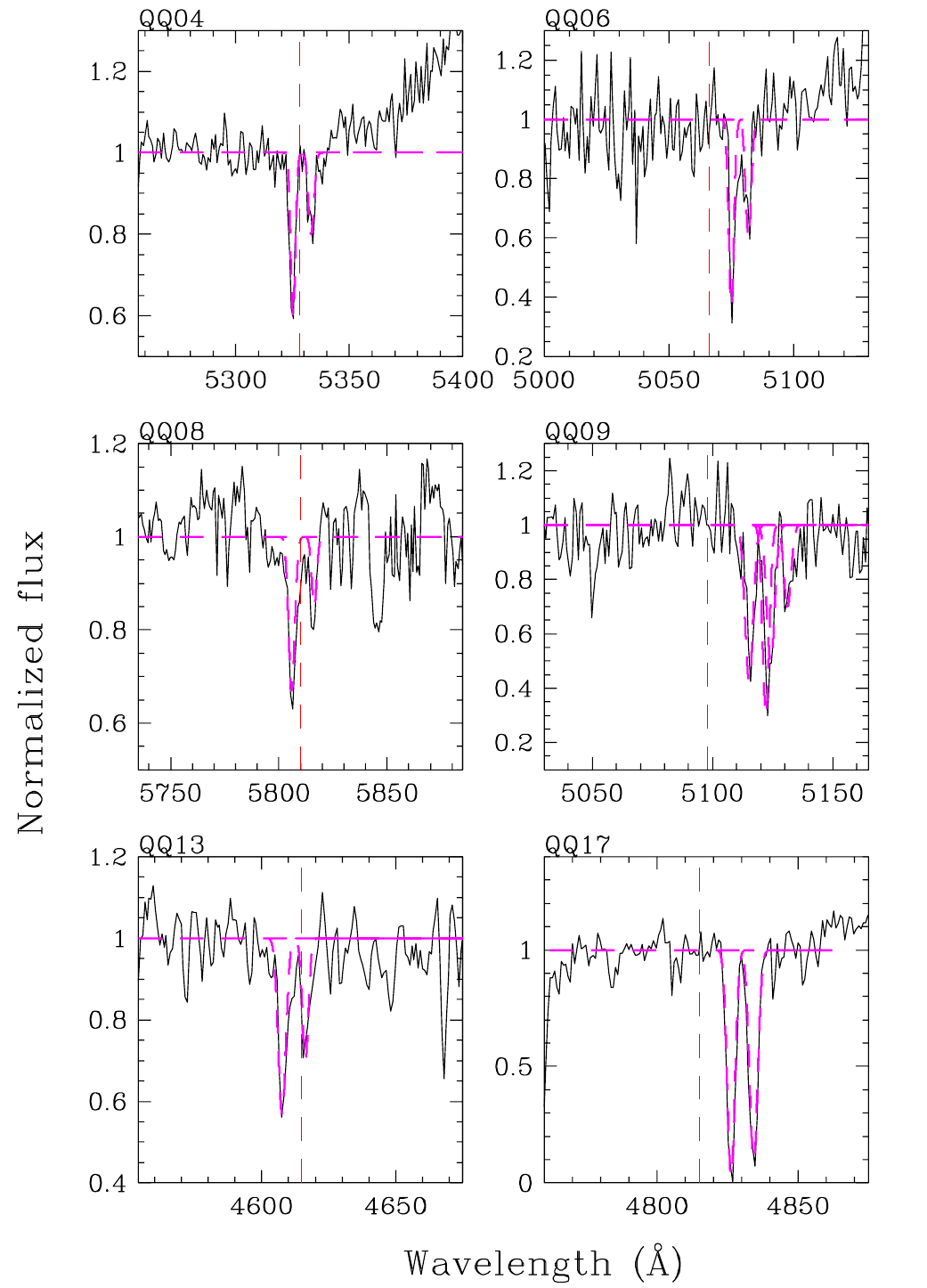}
    \caption{Normalized spectra of background quasars showing the intervening absorption feature identified as C IV 1548$\textrm{\AA}$ and C IV 1550$\textrm{\AA}$ that are close in velocity and projected distance form the foreground quasar (see text). The vertical dashed red line indicate the position of CIV 1548 emission line in the spectrum of the foreground QSO.}
    \label{fig:sampleWin}
\end{figure}

We performed the search of the C IV ($\lambda\lambda 1548-1550)$ absorption doublet in the spectrum of the background QSO for each pair in an interval of wavelengths corresponding to 4000 km s$^{-1}$ centered at the expected position of C IV lines at the redshift of the foreground quasar. In the case of detection we fit the components by the adoption of two gaussian profiles, as illustrated in the boxes of Figure \ref{fig:sampleWin}.  
Furthermore, in order to proper characterise the quality of the data, on each window we measured the minimum detectable Equivalent Width (EW$_{\textrm{min}}$) by following procedure described in \cite*{sbarufatti}. Briefly, we evaluated the EW on bins of the size of the resolution element in various regions of the spectrum excluding telluric structures. We assume as $EW_{min}$ the 2-$\sigma$ deviation from the mean of the average of the distribution of the EWs obtained in each bin. Finally, concerning the LOS velocities of the detected C IV absorptions, we considered that the systems is associated to the foreground object only if $|\Delta V| \leq 500-600$ km s$^{-1}$ rest frame. Results of our procedures are reported in Table \ref{tab:sample}.

\begin{table*}
	\centering
	\caption{Measurement of the C IV absorption in the background QSO of each pair. Identification label (ID), CIV 1548-1551$\lambda$ observed wavelength and equivalent widths (rest),  Doublet Ratio (DR), Velocity difference between the  absorption redshift and the foreground QSO redshift. $\Delta V$ (km s$^{-1}$), Equivalent width minimum detectable on the spectrum ($\textrm{\AA}$).}
	\label{tab:sample}
	\begin{tabular}{llllclll} 
		\hline
		ID & $\lambda_{\textrm{abs}}$(1548)&W (1548) & $\lambda_{\textrm{abs}}$(1551) & W (1551) & DR & $\Delta V$  & EW$_{min}$ \\
		&   \phantom{aaa} [$\textrm{\AA}$] &\phantom{aaa}[$\textrm{\AA}$] &\phantom{aaa} [$\textrm{\AA}$] & [$\textrm{\AA}$] & & [km s$^{-1}$] &\phantom{a} [$\textrm{\AA}$] \\
		\hline

		QQ01B & - & - & - & -& - & - & 0.13 \\
		QQ02B & - &- &- & - & - & - & 0.12  \\
		QQ03B & - & - & -& - & - & - & 0.10  \\
		QQ04B & 5324 & 0.70 $\pm$ 0.20 &5334 & 0.30 $\pm$ 0.10 & 2.30  & -300 & 0.15  \\
		QQ05B &- & - &  - & - & - & - & 0.20  \\
		QQ06B & 5075& 1.30 $\pm$ 0.40 &5084 & 0.60 $\pm$ 0.20 & 2.17  & 500 & 0.15  \\
		QQ07B & - &-&-&  - & - & - & 0.20  \\
		QQ08B & 5805 & 0.50 $\pm$ 0.20 & 5817&0.30 $\pm$ 0.10 & 1.67  & -500 & 0.25  \\
		QQ09B & 5114& 0.60 $\pm$ 0.20 & 5124& 0.30 $\pm$ 0.10 & 2.00  & 600 & 0.18  \\
		QQ10B & - & - &- & - & - & - & 0.20  \\
		QQ11B & - &- &- & - & - & - & 0.16  \\
		QQ12B & - & - &- &-& - & - & 0.10  \\
		QQ13B & 4607 & 0.50 $\pm$ 0.20 &4616 & 0.40 $\pm$ 0.15  & 1.25 & -400 & 0.23  \\
		QQ14B & - &- &- &- & - & - & 0.40  \\
		QQ15B & - & -& -& - & - & - & 0.20  \\
		QQ16B & - &- &-& - & - & - & 0.20  \\
		QQ17B & 4824& 0.50 $\pm0.10 $&4834 & 0.30 $\pm$ 0.05 & 1.67  & 600 & 0.20  \\
		QQ18B &- &- &- & - & - & - & 0.18  \\
		\hline
	\end{tabular}
\end{table*}

\section{Results and Discussion}
\begin{figure*}
	\includegraphics[width=18cm]{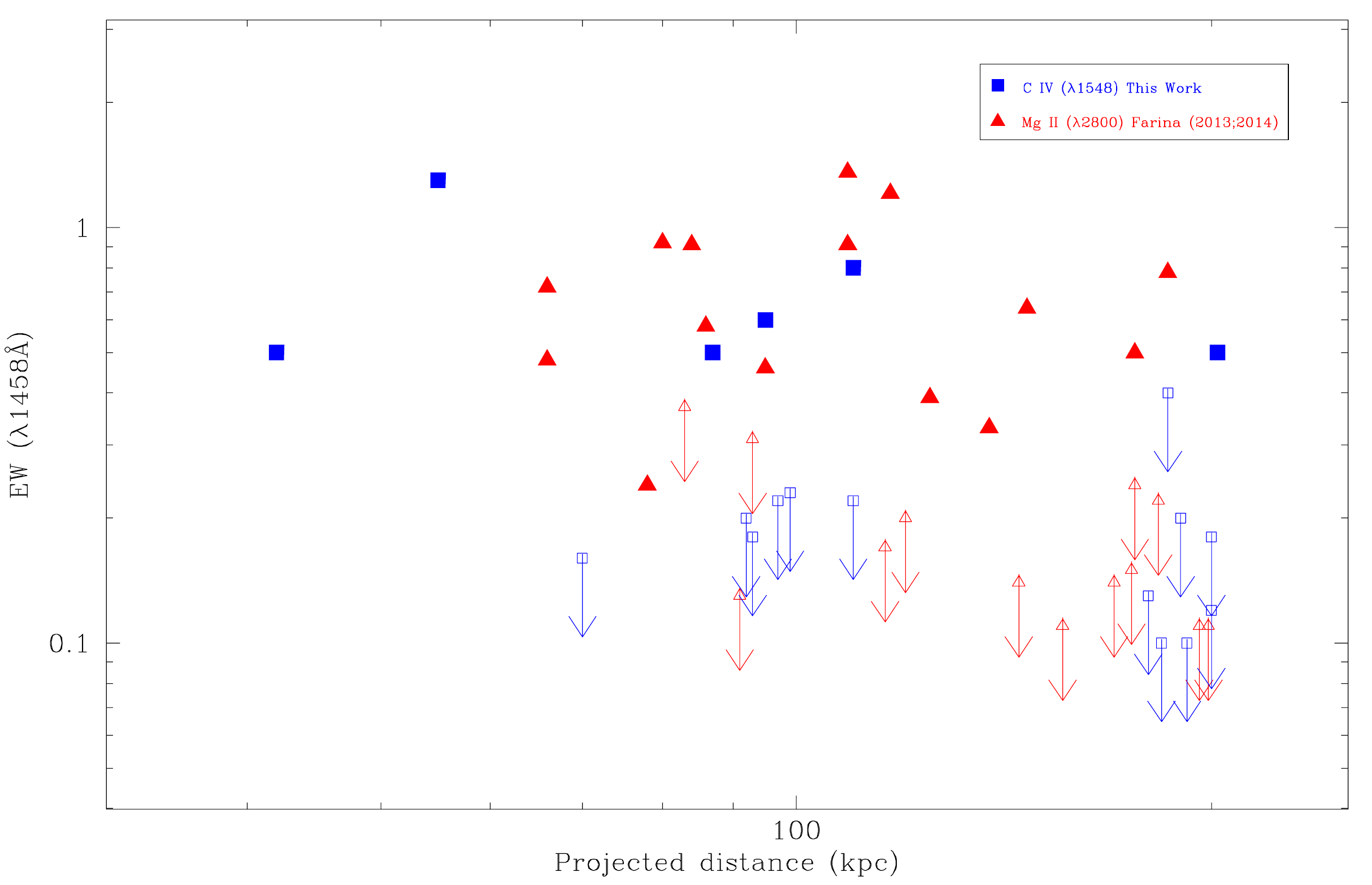}
    \caption{Equivalent width of C IV (filled blue squares) intervening absorption lines as a function of the projected distance 
from the quasar (including the upper limit of QQ10 from Farina et al 2013) . Upper limits are indicated by open squares with arrows (see text). Similar data for Mg II 2800 absorption systems (filled red triangles) and their relative 
upper limits (open triangles with arrows) \citep*{farina13,farina14}.}
    \label{fig:scatter}
\end{figure*}

In 6 out of 18 pairs we detected C IV absorption system associated to the foreground QSO halo (see Table 2 and Figure 2). In one case (QQ09) we have a suggestion of a double C IV systems probably associated to two or more moving clouds belonging to the foreground QSO halos. In this case we deblended the features by fitting the four components adopting gaussian profiles (see box 4 of Figure \ref{fig:sampleWin}). We also note that in QQ10 a C IV absorptions is detected at $\lambda \sim 4800\textrm{\AA}$ in the spectrum of background quasars, but we do not include it in our statistic since the velocity difference is slightly beyond our threshold.
The EW of the detected associated C IV absorption systems together with the upper limits are shown in Figure \ref{fig:scatter} as a function of the projected separation from the foreground QSO. 
In spite of the relatively small statistics, considering both the detections and the upper limits, there is an indication that the absorbing systems decrease in intensity as a function of the distance from the center of the (foreground) quasars and that the absorbing gas becomes more patchy. We performed a Cox Proportional Hazard test including the uppers limits \citep*{isobe} and found that the two quantities (PD and EW) are anti-correlated with a probability of $\sim$ 93\%. This behaviour is qualitatively very similar (see Figure \ref{fig:scatter}) to that of Mg II $\lambda$2800$\textrm{\AA}$ intervening systems \citep*{farina13,farina14} although the average redshift of the objects is somewhat different ($<z> = 1.2$ for Mg II compared with $<z> = 2.1$ for C IV).

\begin{figure}
	\includegraphics[width=8cm]{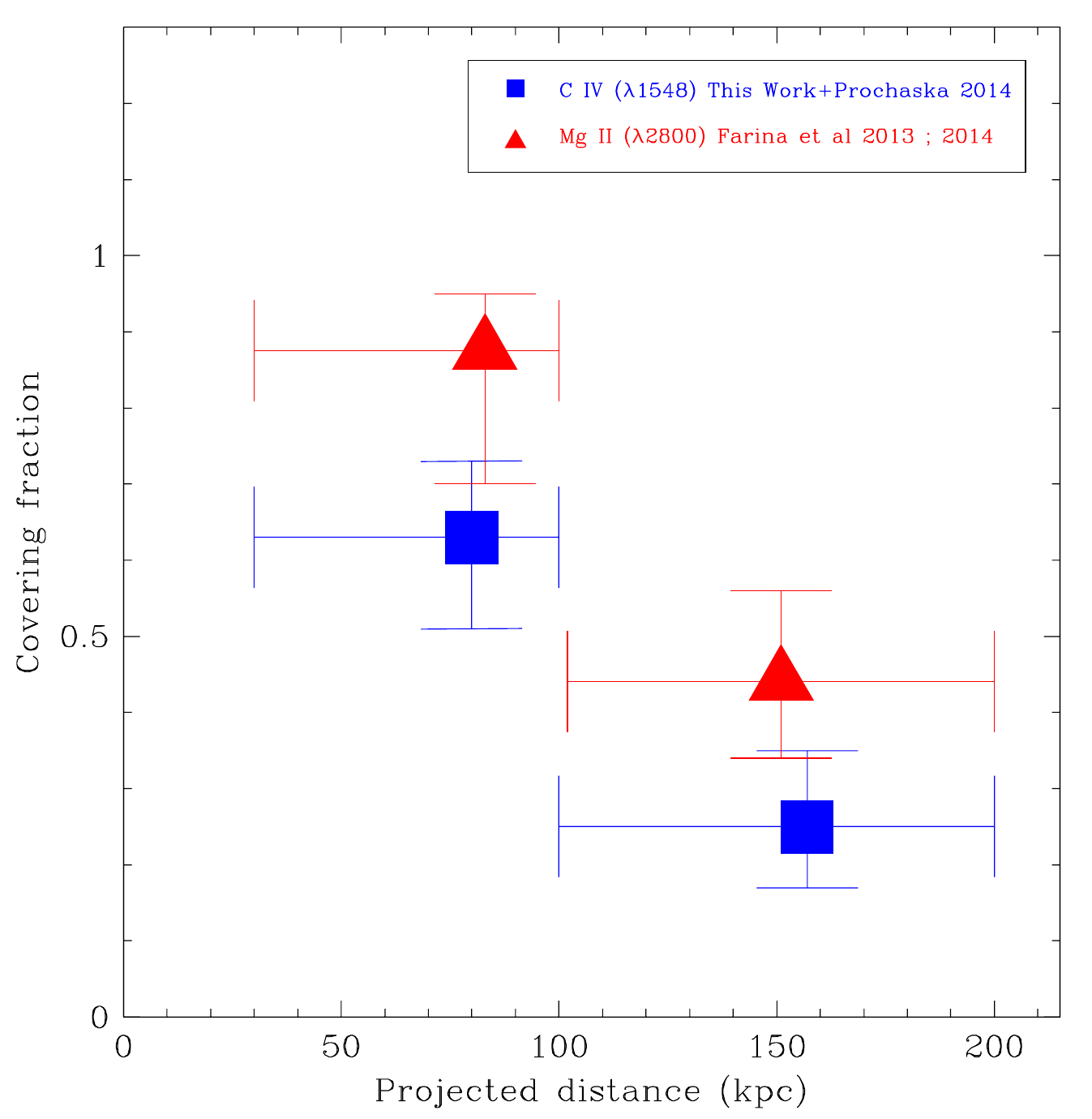}
    \caption{The covering fraction of C IV in QSO host galaxies (filled blue squares) derived from the combined sample (this work plus the dataset by \citep*{prochaska}, see text). For comparison we report the covering fraction of Mg II derived from the dataset considered in \citep*{farina13,farina14} (see text). Vertical error bar are 1-$\sigma$ confidence level at 68\% Wilson score for a binomial distribution.}
    \label{fig:covFrac}
\end{figure}

In order to quantify the patchy structure of the absorbing gaseous halos, we investigate the covering fraction of ($f_c$) of C IV in function of PD. We choose a threshold equivalent width EW$_{th} = 0.30 \textrm{\AA}$, which allows one to consider spectra with EW$_{min} \leq 0.25 \textrm{\AA}$ except for one case (QQ14), and two bins of [0-100] kpc and [100-200] kpc. We define for each bin the $f_c$ as the ratio between detected systems over the total number of pairs in the bin. Since the analysis of the covering fraction is sensitive to binning effect and depends on the adopted EW$_{th}$, we combine our results adopting the same $EW_{th}$ for consistency with those recently drawn in the sample of \cite*{prochaska} yielding 7 extra sources. We find that the covering fraction for C IV is $f_c$ ($\geq$0.30$\AA$) = \textbf{$0.63^{+0.10}_{-0.12}$} for the bin [0-100] kpc while for the case of [100-200] kpc is $f_c$ ($\geq0.30 \textrm{\AA}$) =  $0.25^{+0.10}_{-0.08}$ (see Figure \ref{fig:covFrac}).  Horizontal bars are the bin width while vertical bars are the 1-$\sigma$ uncertainties in the $f_c$ calculated upon the binomial statistics (68\% Wilson score). We note that the $f_c$ of C IV decreases of about a factor of two between the first bin (0-100 kpc) and the second one (100-200 kpc). It is of interest to compare these results with those derived from the covering fraction of the Mg II. We computed, assuming the same bins and EW$_{th}$, the $f_c$ for the Mg II by adopting data presented in \cite*{farina13,farina14} for 26 pairs. We found that $f_c$ ($\geq 0.30\AA$) = $0.86^{+0.10}_{-0.09}$ for the bin [0-100] kpc and $f_c$ ($\geq 0.30 \textrm{\AA}$) = $0.45^{+0.13}_{-0.10}$ for [100-200] kpc (see Figure \ref{fig:covFrac}).

Both for Mg II and CIV species, the covering fraction of the absorbing material is halved from the region (< 100 kpc) close to the center of the host galaxy to the immediate outer region (100-200 kpc). There is a suggestion that the covering fraction of CIV absorbers is systematically smaller than that of MgII. This behaviour could be related due to different ionisation energies of the two species and/or due to chemical abundances. We note that, although the statistics is small, our finding is also consistent with results based on C II and C IV for a sample of 60 quasar pairs \citep{prochaska}. 

\section*{Acknowledgements}
EPF acknowledges funding through the ERC grant `Cosmic Dawn'.

%
%
%




\begin{thebibliography}{99}
\bibitem[Adelberger et al.(2005)]{adel} Adelberger, K.~L., Shapley, A.~E., Steidel, C.~C., et al.\ 2005, \apj, 629, 636 
\bibitem[Bahcall \& Spitzer(1969)]{bs} Bahcall, J.~N., \& Spitzer, L., Jr.\ 1969, \apjl, 156, L63 
\bibitem[Cantalupo et al.(2014)]{cantalupo} Cantalupo, S., Arrigoni-Battaia, F., Prochaska, J.~X., Hennawi, J.~F., \& Madau, P.\ 2014, \nat,506, 63 
\bibitem[Cardelli et al.(1989)]{cardelli} Cardelli, J.~A., Clayton, G.~C., \& Mathis, J.~S.\ 1989, \apj, 345, 245 
\bibitem[Cepa et al.(2003)]{cepa} Cepa, J., Aguiar-Gonzalez, 
M., Bland-Hawthorn, J., et al.\ 2003, \procspie, 4841, 1739 
\bibitem[Chen et al.(2001)]{chen} Chen, H.-W., Lanzetta, 
K.~M., Webb, J.~K., \& Barcons, X.\ 2001, \apj, 559, 654 
\bibitem[Chen \& Tinker(2008)]{chen08} Chen, H.-W., \& Tinker, J.~L.\ 2008, \apj, 687, 745 
\bibitem[Churchill et al.(2005)]{churchill} Churchill, C., 
Steidel, C., \& Kacprzak, G.\ 2005, Extra-Planar Gas, 331, 387 
\bibitem[Churchill et al.(2013)]{churchill13} Churchill, C.~W., 
Nielsen, N.~M., Kacprzak, G.~G., 
\& Trujillo-Gomez, S.\ 2013, \apjl, 763, L42 
\bibitem[Cortese et 
al.(2006)]{cortese} Cortese, L., Gavazzi, G., Boselli, A., et al.\ 2006, \aap, 453, 847 
\bibitem[Di Matteo et al.(2005)]{dimatteo} Di Matteo, T., 
Springel, V., \& Hernquist, L.\ 2005, \nat, 433, 604 
\bibitem[Farina et al.(2013)]{farina13} Farina, E.~P., Falomo, R., Decarli, R., Treves, A., \& Kotilainen, J.~K.\ 2013, \mnras, 429, 1267 
\bibitem[Farina et al.(2014)]{farina14} Farina, E.~P., Falomo, R., Scarpa, R., et al.\ 2014, \mnras, 441, 886 
\bibitem[Hennawi et al.(2006)]{hennawi} Hennawi, J.~F., 
Prochaska, J.~X., Burles, S., et al.\ 2006, \apj, 651, 61 
\bibitem[Hennawi 
\& Prochaska(2007)]{hp07} Hennawi, J.~F., \& Prochaska, J.~X.\ 2007, \apj, 655, 735 
\bibitem[Hennawi 
\& Prochaska(2013)]{hp13} Hennawi, J.~F., \& Prochaska, J.~X.\ 2013, \apj, 766, 58 
\bibitem[Hennawi et al.(2015)]{hennawi15} Hennawi, J.~F., 
Prochaska, J.~X., Cantalupo, S., 
\& Arrigoni-Battaia, F.\ 2015, Science, 348, 779 
\bibitem[Isobe et al.(1986)]{isobe} Isobe, T., Feigelson, 
E.~D., \& Nelson, P.~I.\ 1986, \apj, 306, 490 
\bibitem[Lanzetta et al.(1995)]{lanz} Lanzetta, K.~M., 
Bowen, D.~V., Tytler, D., \& Webb, J.~K.\ 1995, \apj, 442, 538 
\bibitem[Martin et al.(2014)]{martin} Martin, D.~C., Chang, 
D., Matuszewski, M., et al.\ 2014, \apj, 786, 106 
\bibitem[M{\'e}nard et al.(2011)]{menard} M{\'e}nard, B., 
Wild, V., Nestor, D., et al.\ 2011, \mnras, 417, 801 
\bibitem[Nestor et al.(2011)]{nestor} Nestor, D.~B., Johnson, 
B.~D., Wild, V., et al.\ 2011, \mnras, 412, 1559 
\bibitem[Nielsen et al. (2013a)]{mag1} Nielsen, N.M., Churchill, C.~W., Kacprzak, G.~G., \& Murphy, M.~T.\ 2013, \apj, 776, 114 
\bibitem[Nielsen et al. (2013b)]{mag2} Nielsen, N. M., Churchill, C.~W., \& Kacprzak, G.~G.\ 2013, \apj, 776, 115 
\bibitem[Paris et al.(2012)]{paris} Paris, I., Petitjean, P., Aubourg, {\'E}., et al.\ 2012, \aap, 548, A66 
\bibitem[Prochaska et al.(2014)]{prochaska} Prochaska, J.~X., 
Lau, M.~W., \& Hennawi, J.~F.\ 2014, \apj, 796, 140 
\bibitem[Prochter et al.(2006)]{prochter} Prochter, G.~E., 
Prochaska, J.~X., \& Burles, S.~M.\ 2006, \apj, 639, 766 
\bibitem[Sbarufatti et al.(2006)]{sbarufatti} Sbarufatti, B., 
Treves, A., Falomo, R., et al.\ 2006, \aj, 132, 1 
\bibitem[Schlegel et al.(1998)]{sch} Schlegel, D.~J., 
Finkbeiner, D.~P., \& Davis, M.\ 1998, \apj, 500, 525 
\bibitem[Steidel et al.(2010)]{steidel} Steidel, C.~C., Erb, D.~K., Shapley, A.~E., et al.\ 2010, \apj, 717, 289 
\bibitem[Shen 
\& M{\'e}nard(2012)]{shen} Shen, Y., \& M{\'e}nard, B.\ 2012, \apj, 748, 131 
\bibitem[Sulentic et al.(2001)]{sulentic} Sulentic, J.~W., 
Rosado, M., Dultzin-Hacyan, D., et al.\ 2001, \aj, 122, 2993 
\bibitem[Zibetti et al.(2007)]{zibetti} Zibetti, S., 
M{\'e}nard, B., Nestor, D.~B., et al.\ 2007, \apj, 658, 161 

\end{thebibliography}



\clearpage

\section*{QSO pairs spectra for the referee documentation}
%
\begin{figure*}
	\includegraphics[width=18cm]{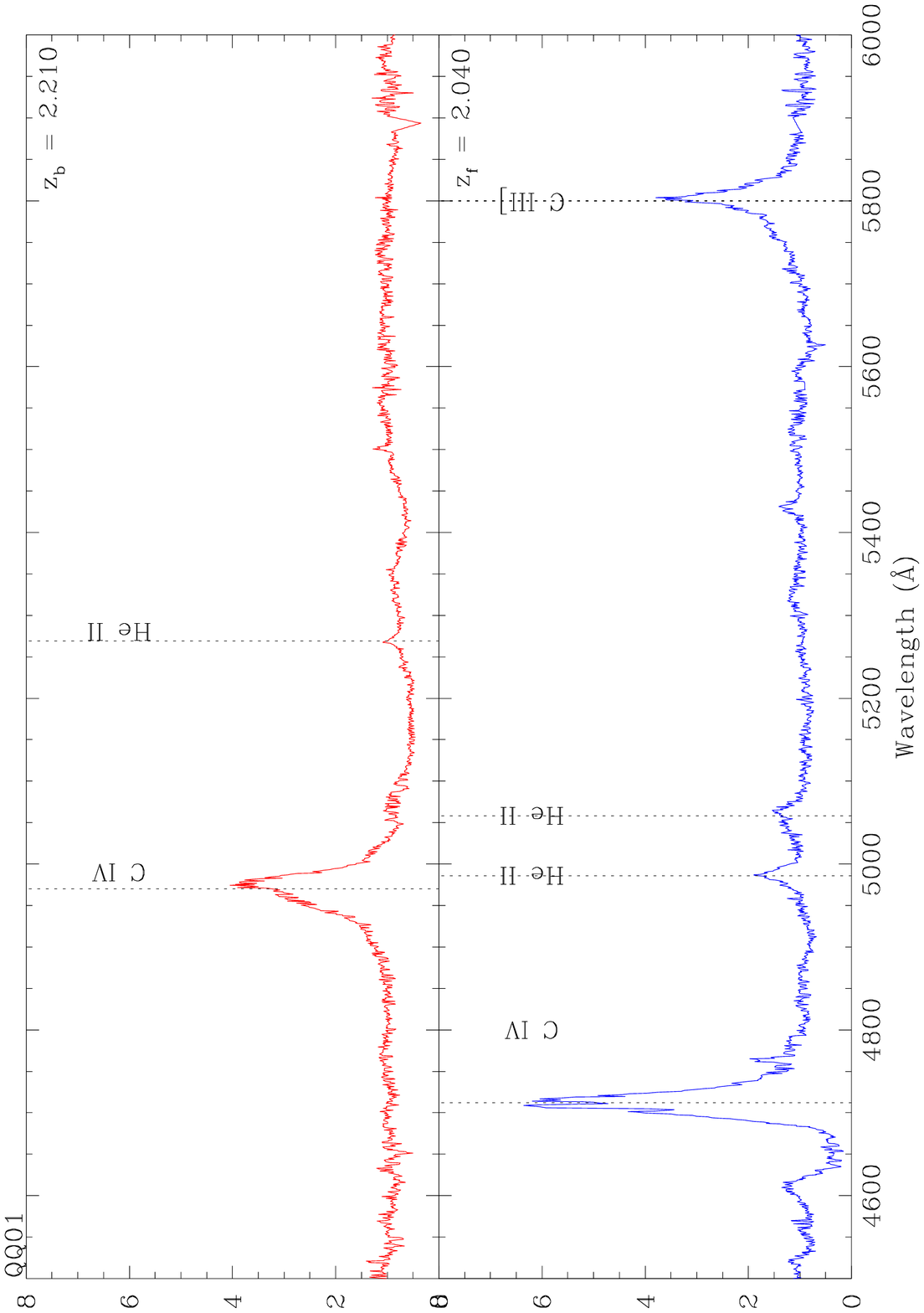}
    \caption{Spectra of pair QQ01.}
    \label{fig:qq01pair}
\end{figure*}
\begin{figure*}
	\includegraphics[width=18cm]{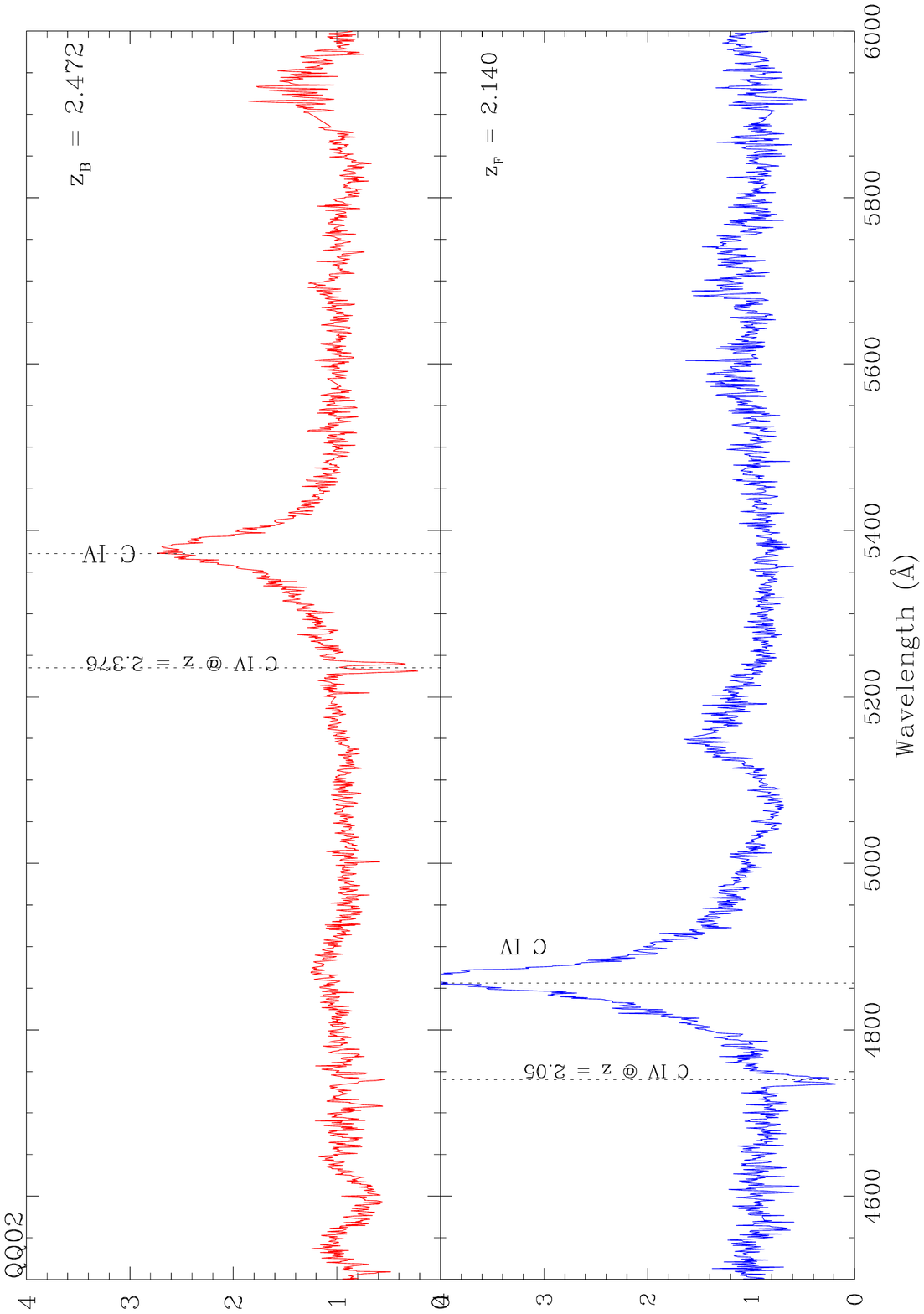}
    \caption{Spectra of pair QQ02.}
    \label{fig:qq01pair}
\end{figure*}

\begin{figure*}
	\includegraphics[width=18cm]{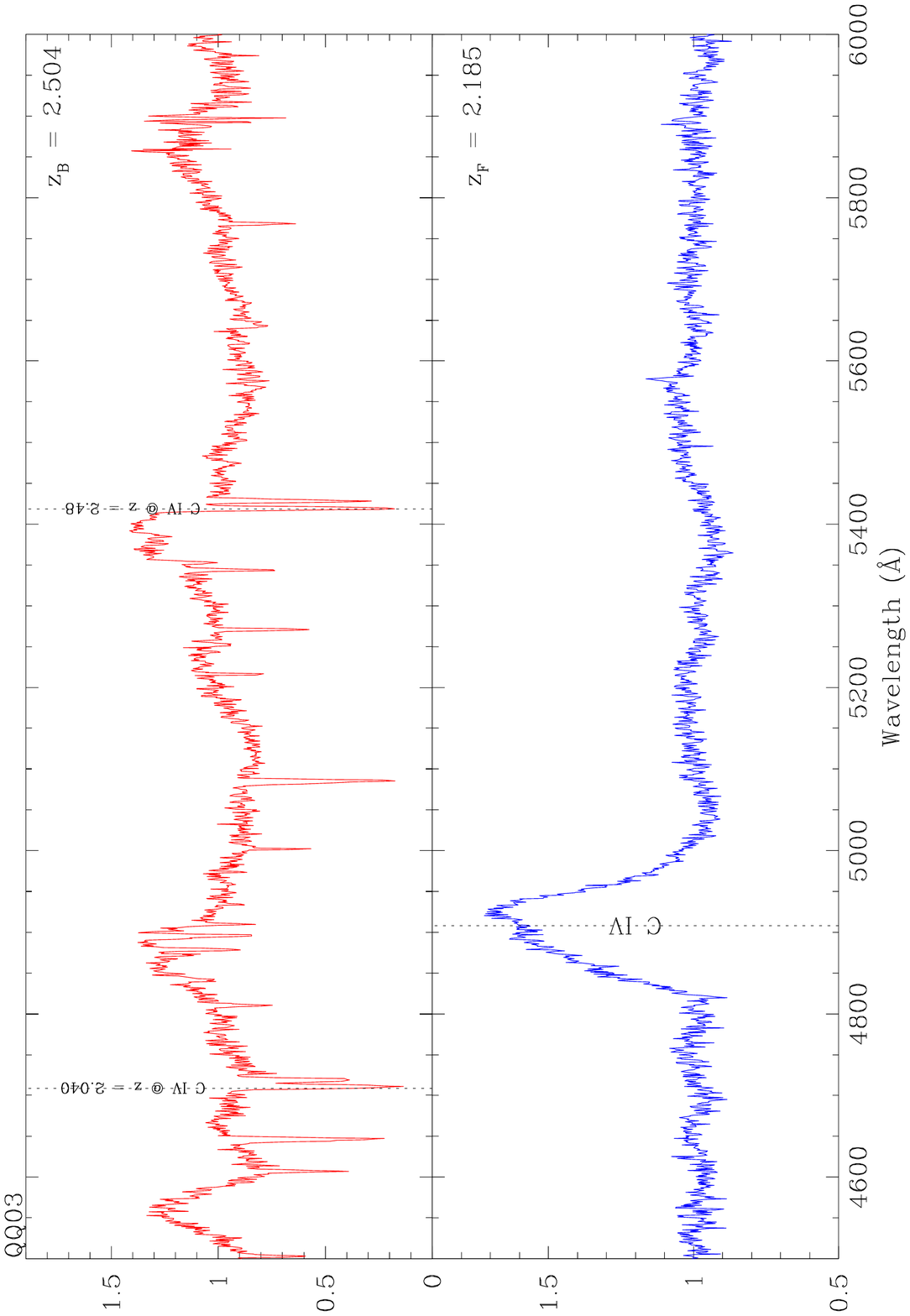}
    \caption{Spectra of pair QQ03.}
    \label{fig:qq01pair}
\end{figure*}

\begin{figure*}
	\includegraphics[width=18cm]{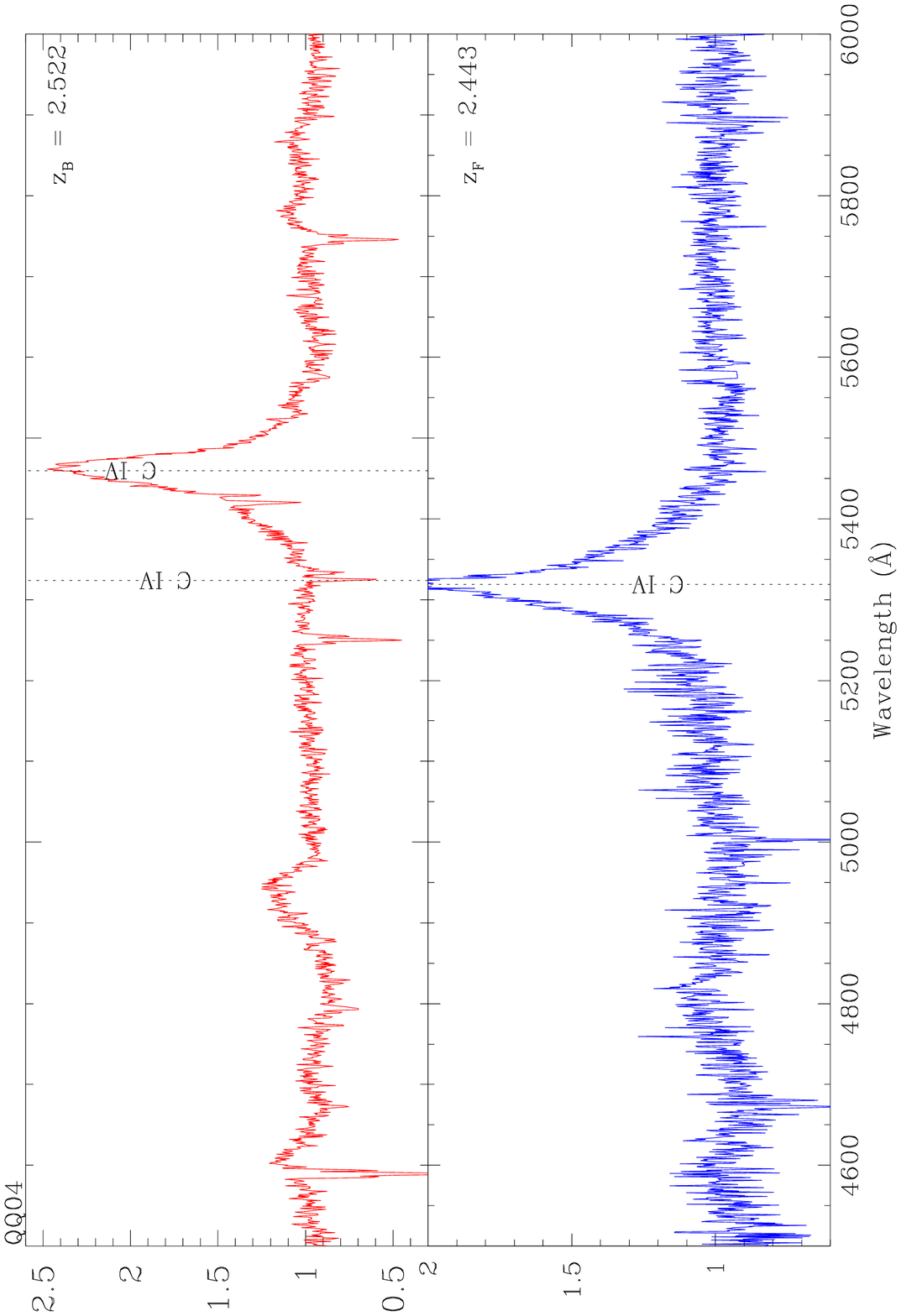}
    \caption{Spectra of pair QQ04.}
    \label{fig:qq01pair}
\end{figure*}

\begin{figure*}
	\includegraphics[width=18cm]{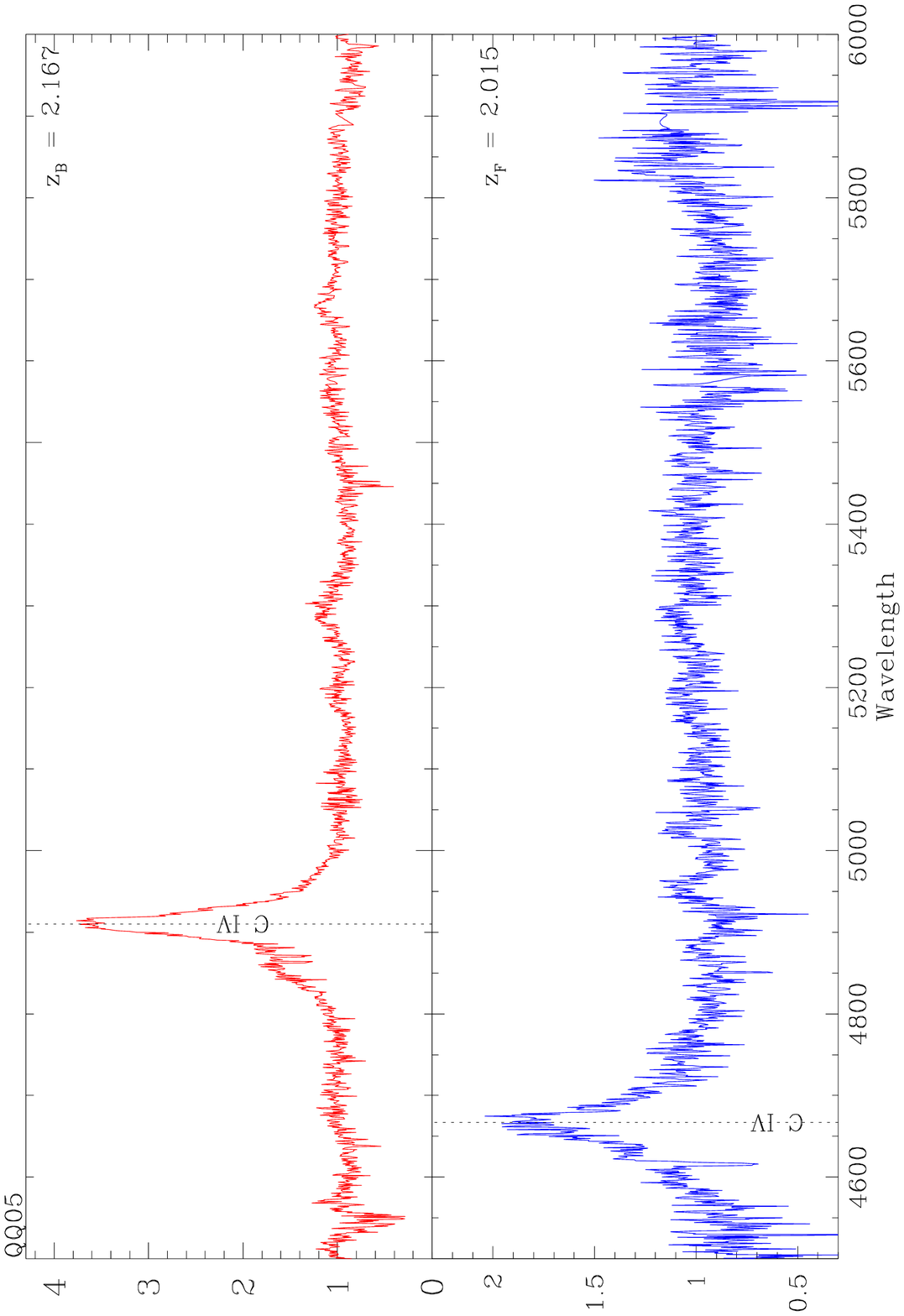}
    \caption{Spectra of pair QQ05.}
    \label{fig:qq01pair}
\end{figure*}

\begin{figure*}
	\includegraphics[width=18cm]{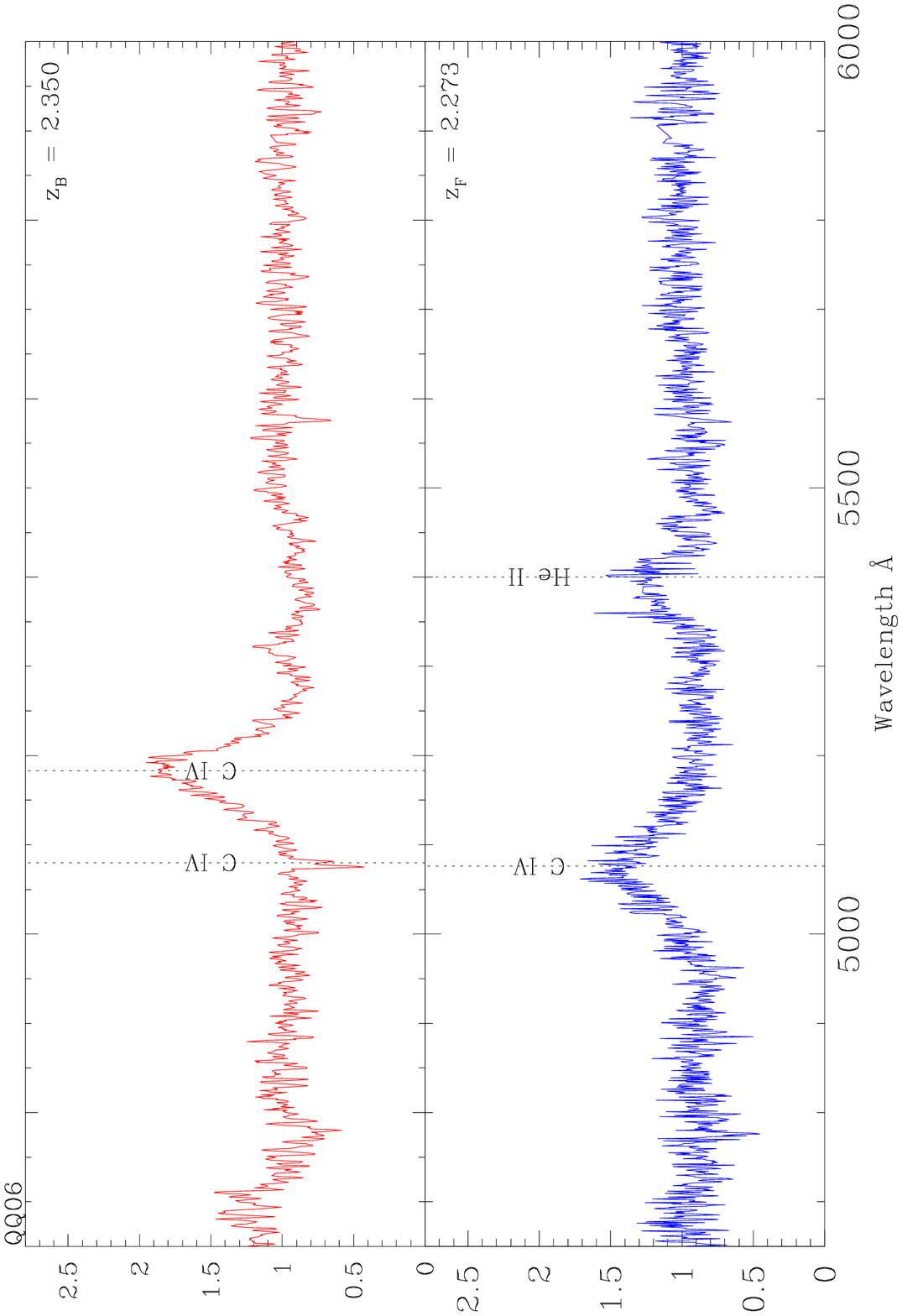}
    \caption{Spectra of pair QQ06.}
    \label{fig:qq01pair}
\end{figure*}

\begin{figure*}
	\includegraphics[width=18cm]{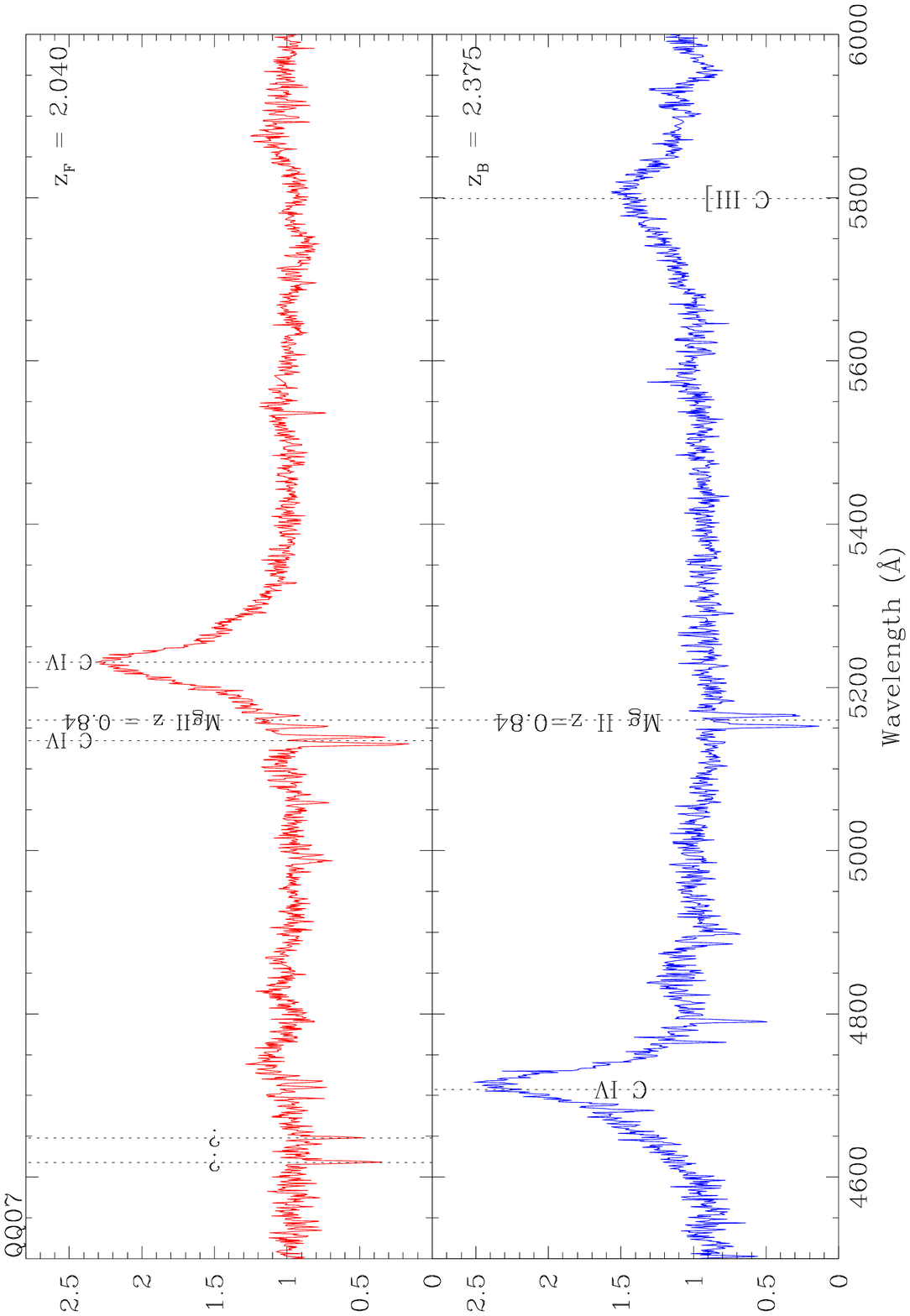}
    \caption{Spectra of pair QQ07.}
    \label{fig:qq01pair}
\end{figure*}

\begin{figure*}
	\includegraphics[width=18cm]{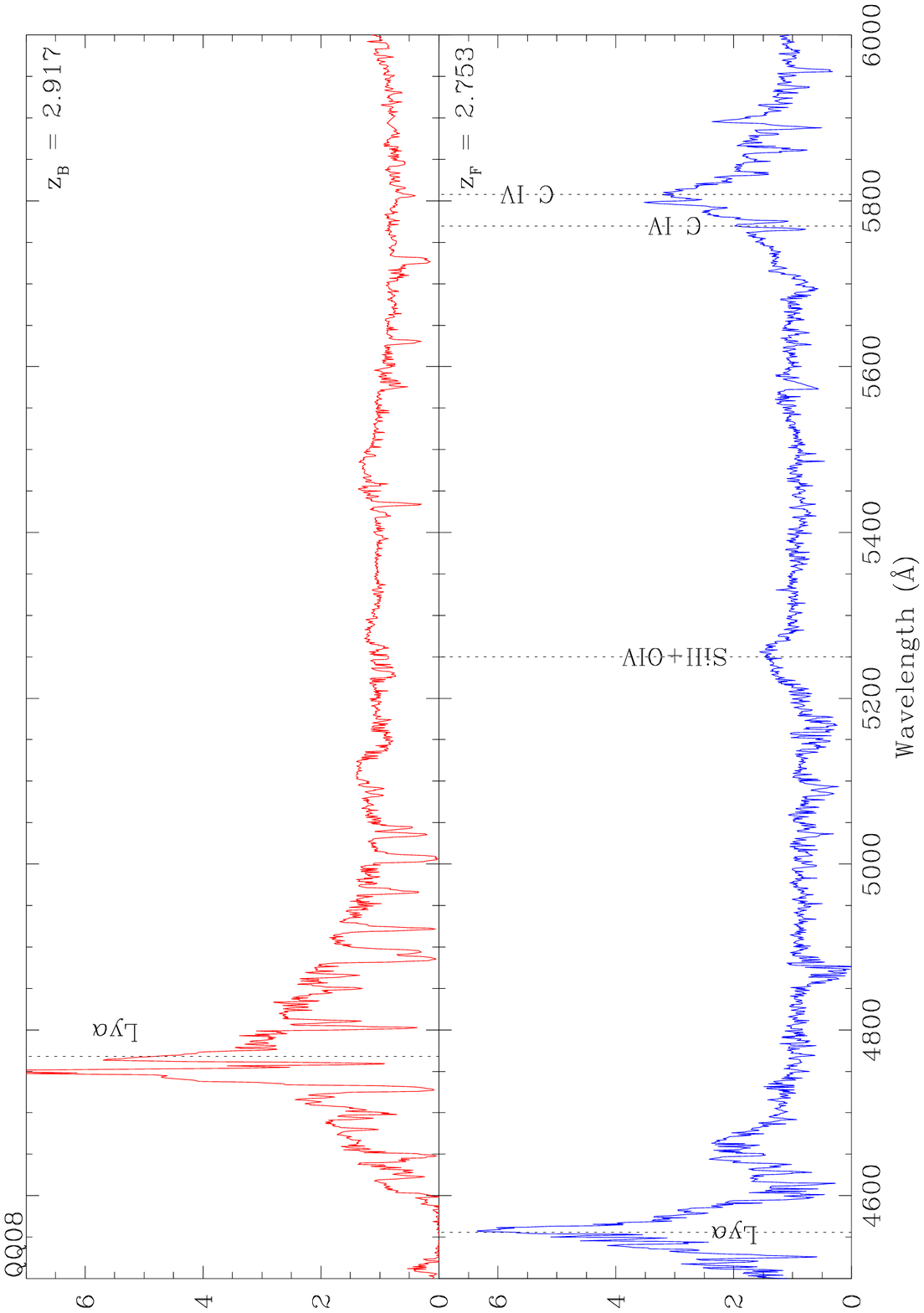}
    \caption{Spectra of pair QQ08.}
    \label{fig:qq01pair}
\end{figure*}

\begin{figure*}
	\includegraphics[width=18cm]{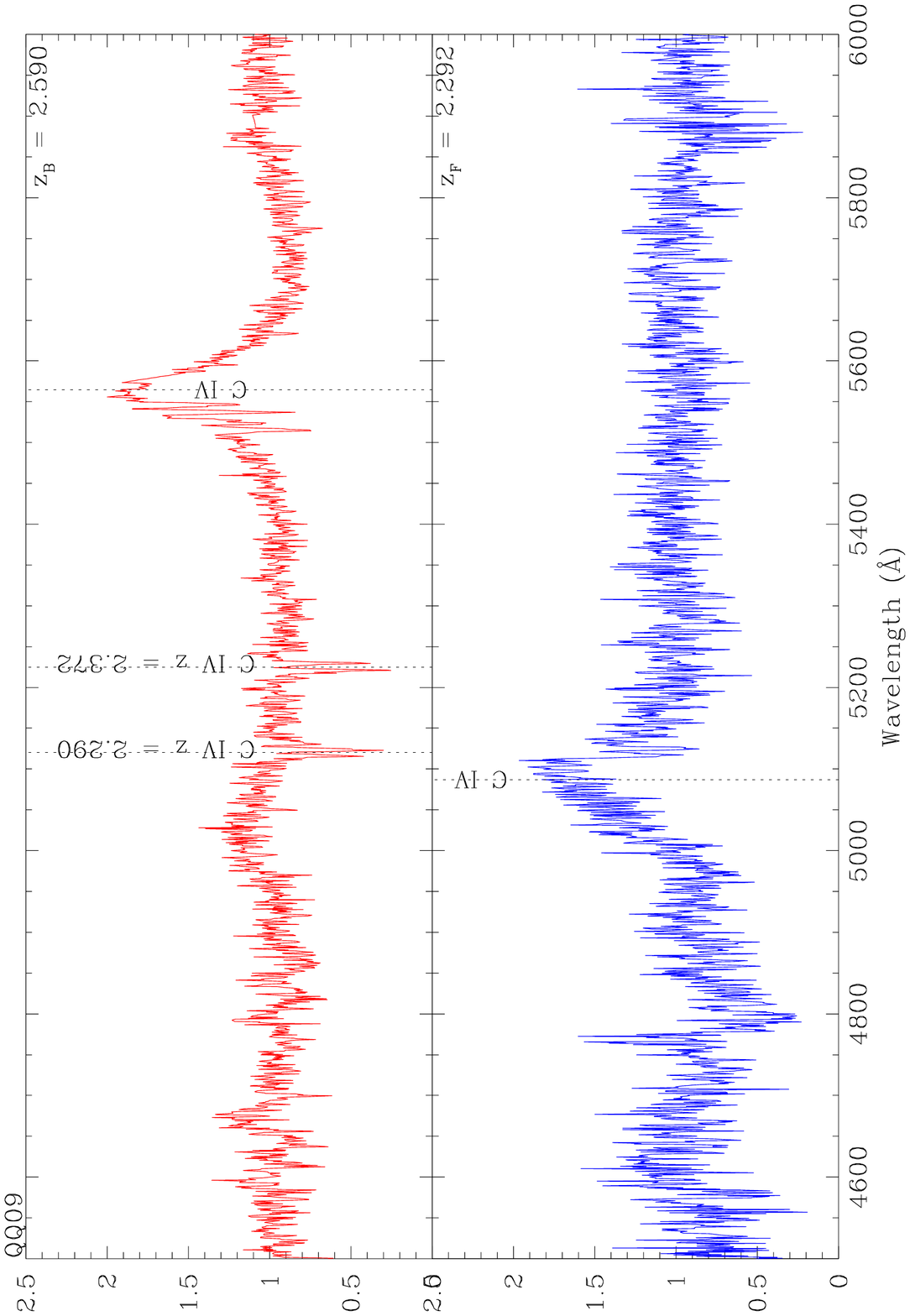}
    \caption{Spectra of pair QQ09.}
    \label{fig:qq01pair}
\end{figure*}

\begin{figure*}
	\includegraphics[width=18cm]{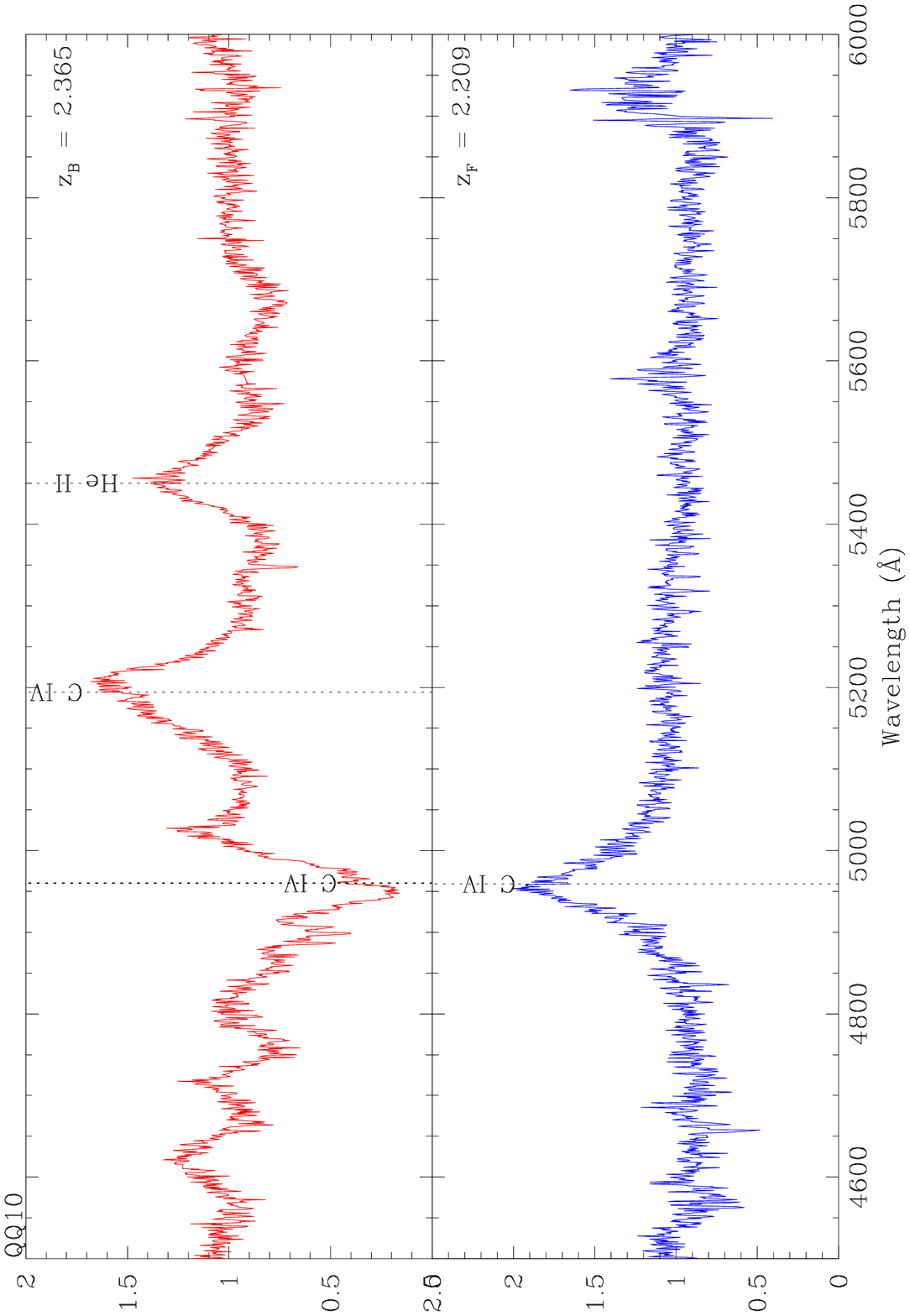}
    \caption{Spectra of pair QQ10.}
    \label{fig:qq01pair}
\end{figure*}

\begin{figure*}
	\includegraphics[width=18cm]{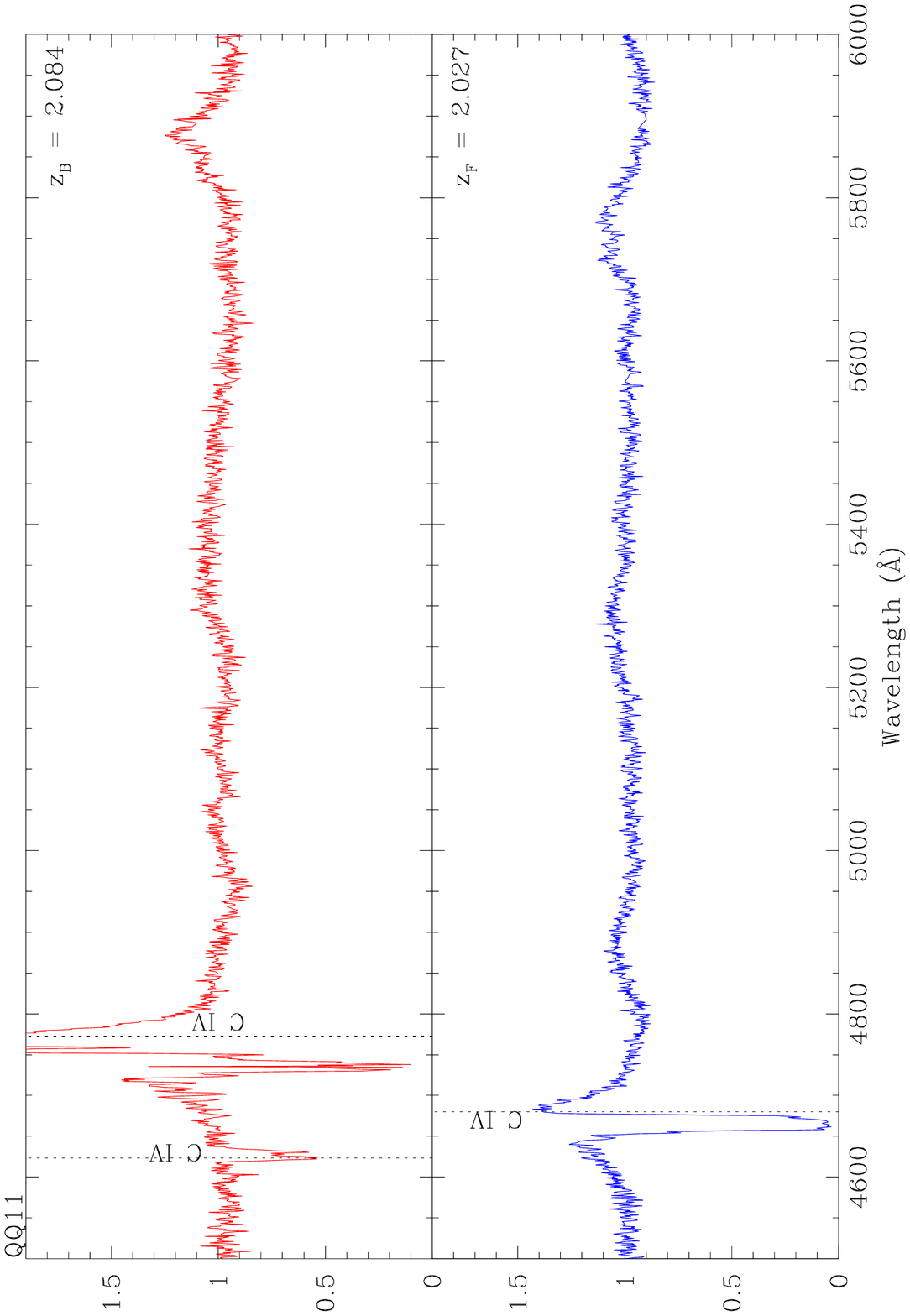}
    \caption{Spectra of pair QQ11.}
    \label{fig:qq01pair}
\end{figure*}

\begin{figure*}
	\includegraphics[width=18cm]{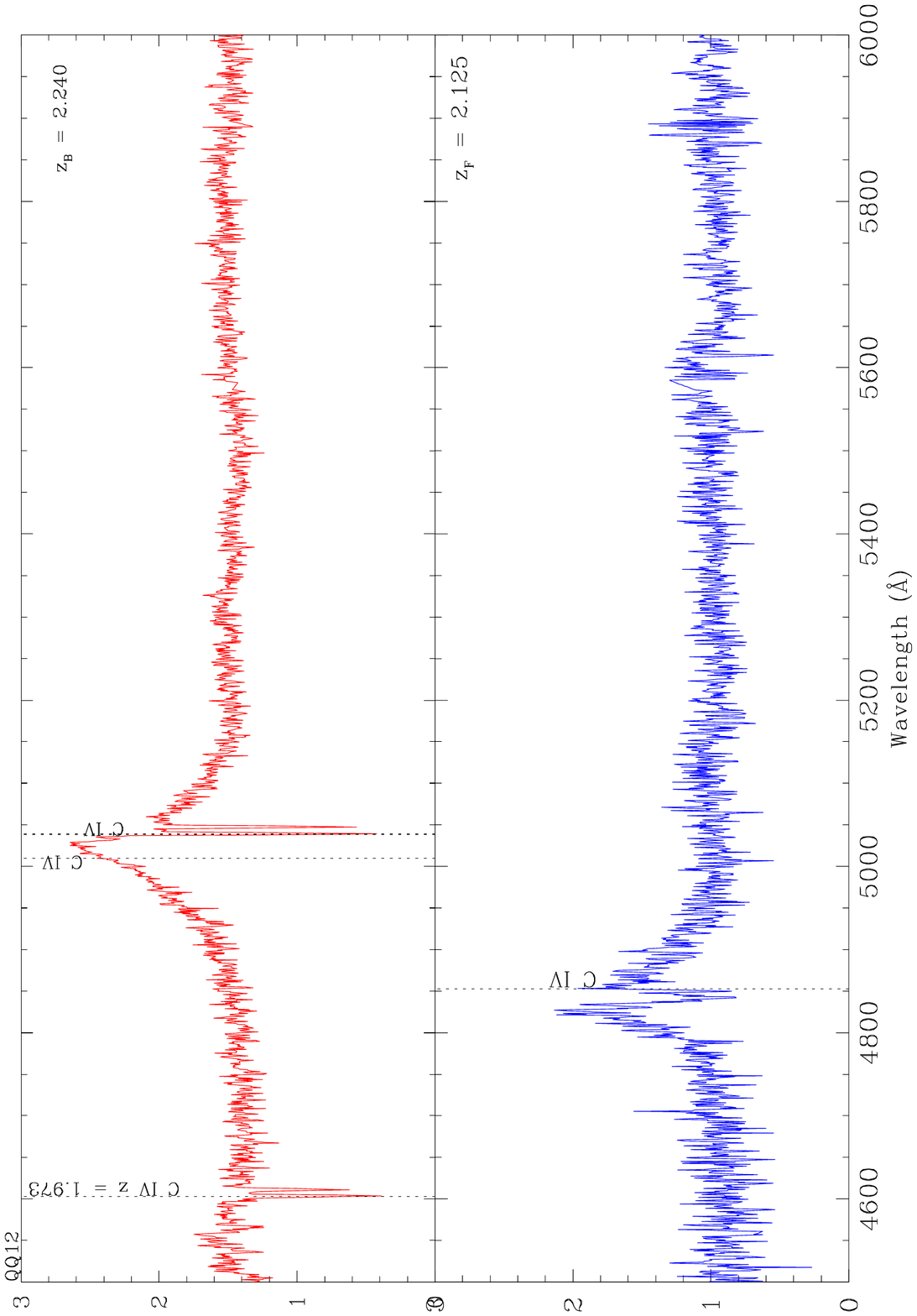}
    \caption{Spectra of pair QQ12.}
    \label{fig:qq01pair}
\end{figure*}

\clearpage

\begin{figure*}
	\includegraphics[width=18cm]{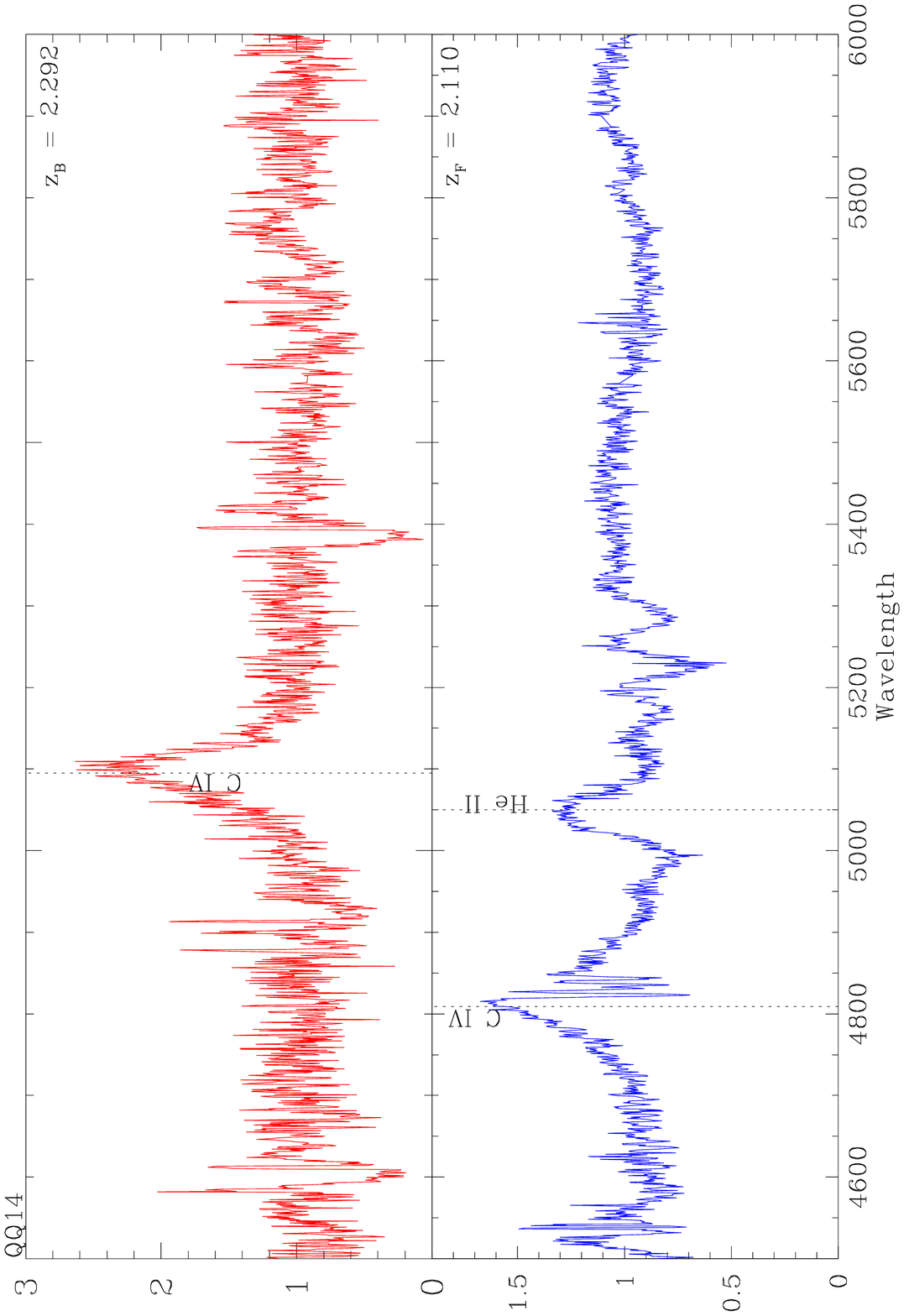}
    \caption{Spectra of pair QQ14.}
    \label{fig:qq01pair}
\end{figure*}
\begin{figure*}
	\includegraphics[width=18cm]{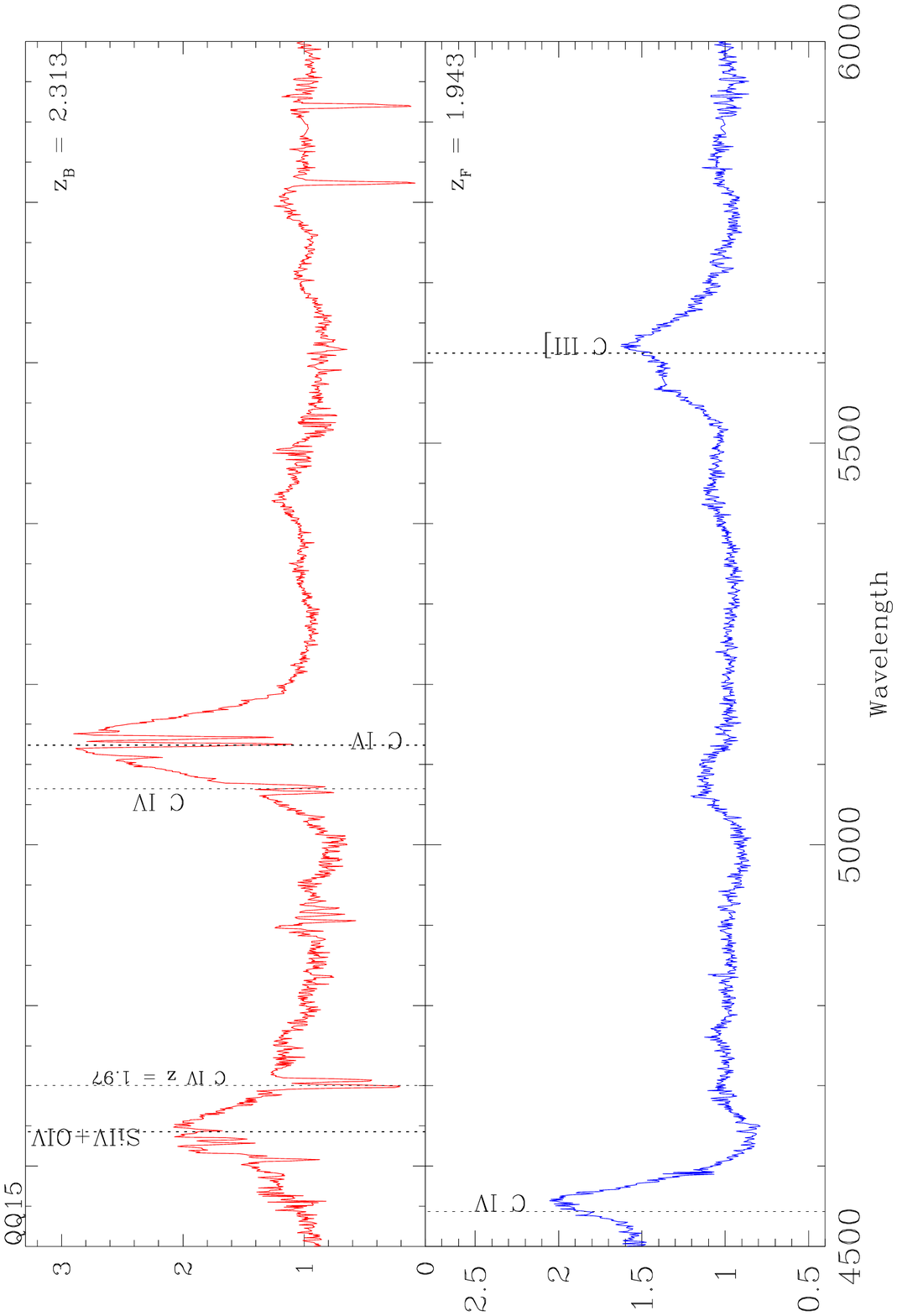}
    \caption{Spectra of pair QQ15.}
    \label{fig:qq01pair}
\end{figure*}

\begin{figure*}
	\includegraphics[width=18cm]{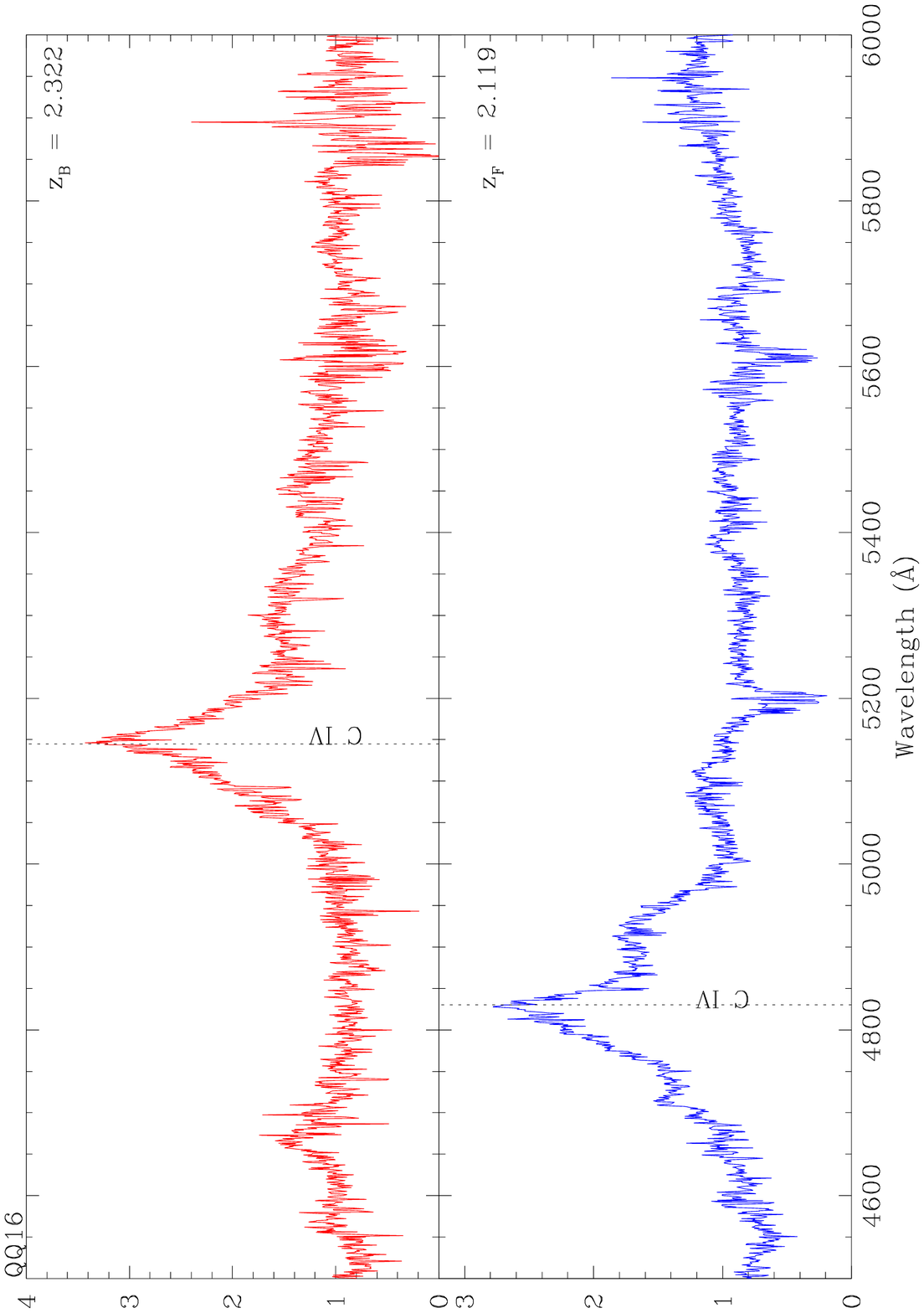}
    \caption{Spectra of pair QQ16.}
    \label{fig:qq01pair}
\end{figure*}

\begin{figure*}
	\includegraphics[width=18cm]{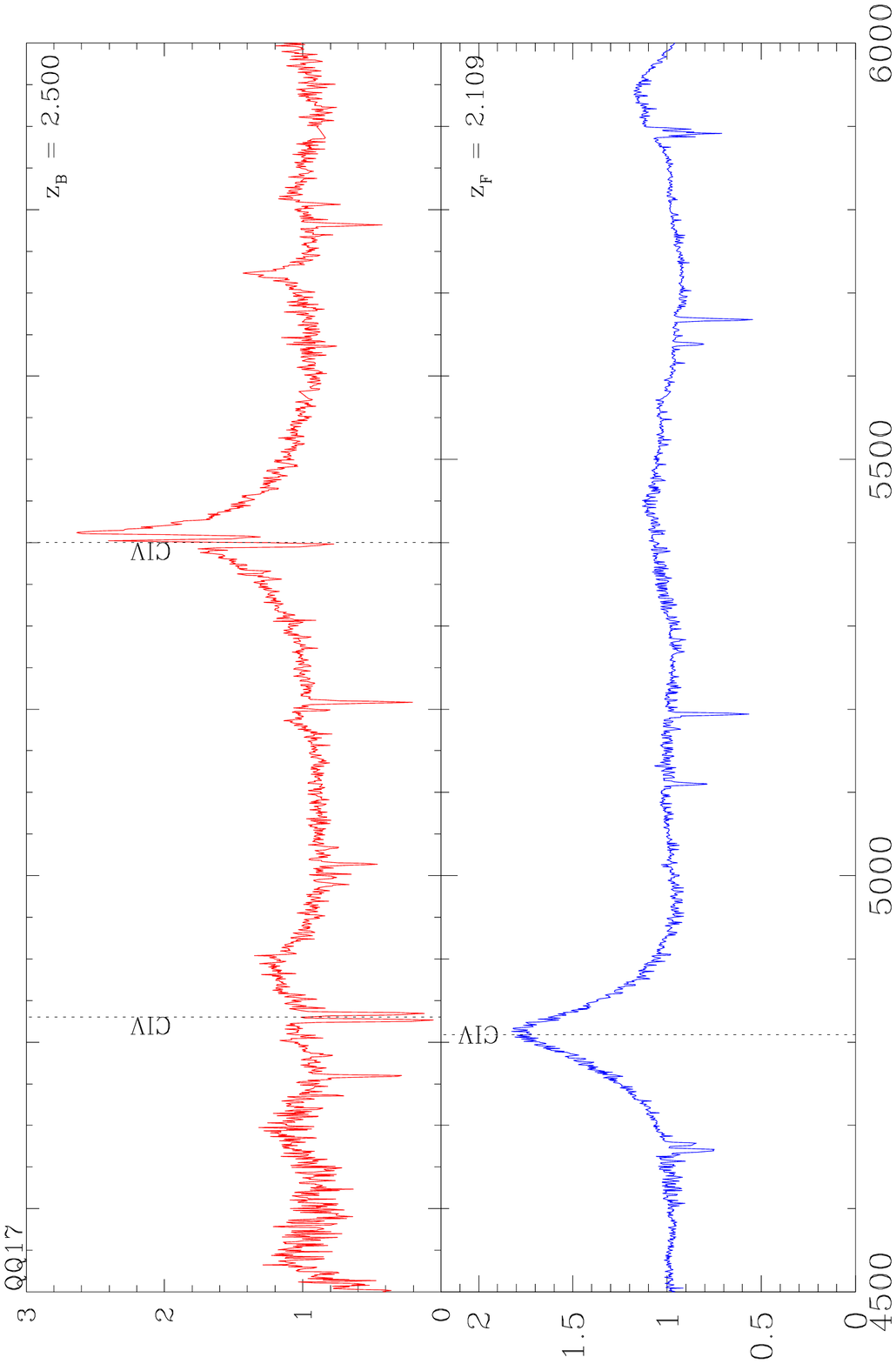}
    \caption{Spectra of pair QQ17.}
    \label{fig:qq01pair}
\end{figure*}

\begin{figure*}
	\includegraphics[width=18cm]{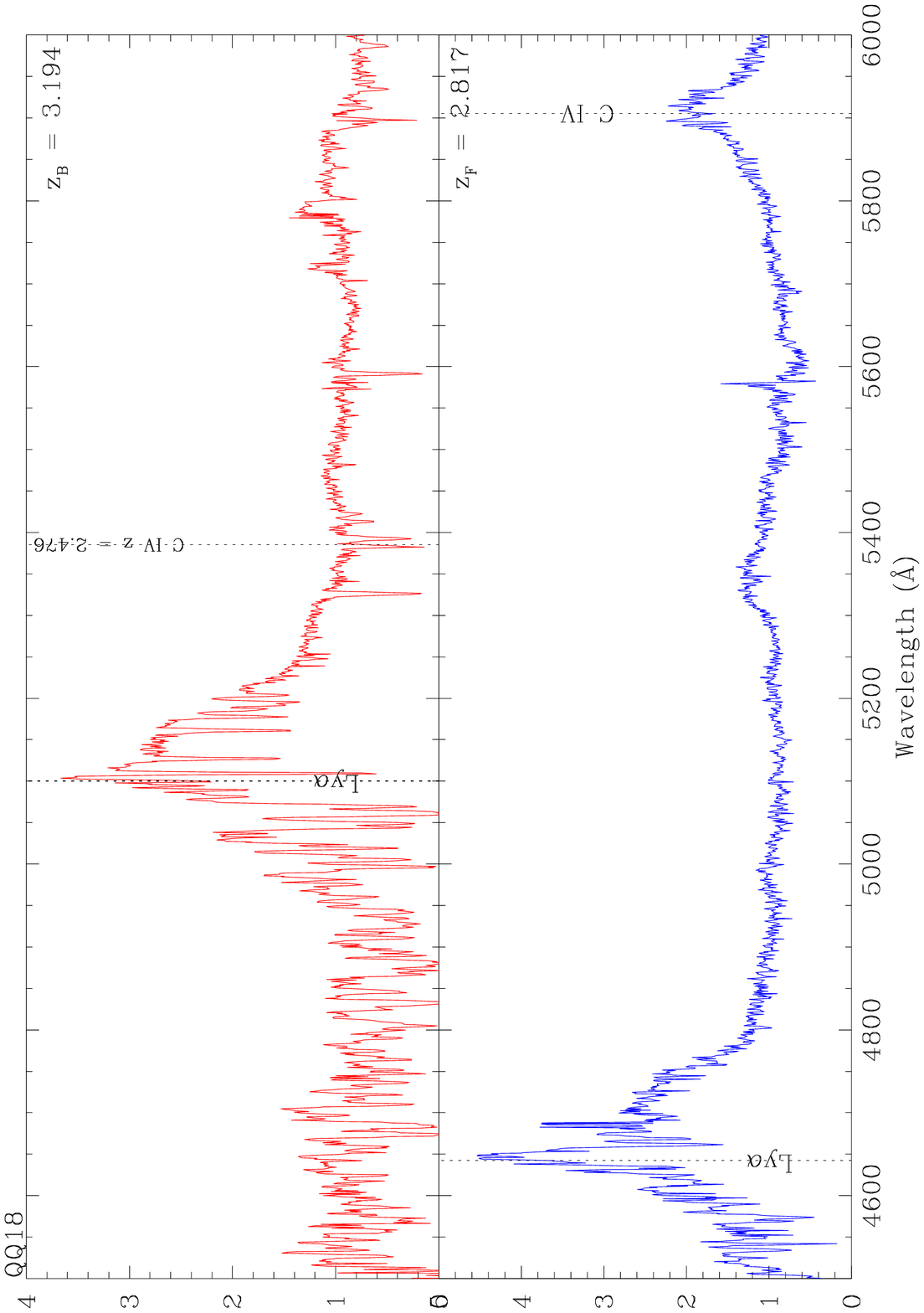}
    \caption{Spectra of pair QQ18.}
    \label{fig:qq01pair}
\end{figure*}
\bsp	
\label{lastpage}
\end{document}